\documentclass[twocolumn,superscriptaddress, showpacs] {revtex4}
\usepackage{graphicx}
\usepackage{amsmath}
\usepackage{subfigure}
\usepackage{dsfont}
\usepackage{amssymb}
\usepackage{float}

\begin{document}

\title[Short Title]{Dissipative stabilization of quantum-feedback-based multipartite entanglement with Rydberg atoms}
\author{Xiao-Qiang Shao\footnote{Corresponding author: shaoxq644@nenu.edu.cn}}
\affiliation{Center for Quantum Sciences and School of Physics, Northeast Normal University, Changchun 130024, People¡¯s Republic of China,
and Center for Advanced Optoelectronic Functional Materials Research and Key Laboratory for UV Light-Emitting Materials and
Technology of Ministry of Education, Northeast Normal University, Changchun 130024, People¡¯s Republic of China}

\author{Jin-Hui Wu}
\affiliation{Center for Quantum Sciences and School of Physics, Northeast Normal University, Changchun 130024, People¡¯s Republic of China,
and Center for Advanced Optoelectronic Functional Materials Research and Key Laboratory for UV Light-Emitting Materials and
Technology of Ministry of Education, Northeast Normal University, Changchun 130024, People¡¯s Republic of China}

\author{Xue-Xi Yi}
\affiliation{Center for Quantum Sciences and School of Physics, Northeast Normal University, Changchun 130024, People¡¯s Republic of China,
and Center for Advanced Optoelectronic Functional Materials Research and Key Laboratory for UV Light-Emitting Materials and
Technology of Ministry of Education, Northeast Normal University, Changchun 130024, People¡¯s Republic of China}

\begin{abstract}
A quantum-feedback-based scheme is proposed for generating multipartite entanglements of Rydberg atoms in a dissipative optical cavity. The Rydberg blockade mechanism efficiently prevents double excitations of the system, which is further exploited to speed up the stabilization of an entangled state with a single Rydberg state excitation. The corresponding feedback operations are greatly simplified, since only one regular atom needs to be controlled during the whole process, irrespective of the number of particles. The form of entangled state is also adjustable via regulating the Rabi frequencies of driving fields. Moreover, a relatively long-life time of the high-lying Rydberg level guarantees a high fidelity in a realistic situation.
\end{abstract}
 \maketitle \maketitle
\section{Introduction}
Quantum entanglement, formally proposed by Ervin Schr\"{o}dinger, is defined to describe a strongly correlated system constituted by pairs or groups of particles \cite{sc}. This kind of correlation is so peculiar that a measurement made on either of the particles apparently collapses the state of system instantaneously, even when the particles are separated by a large distance. Although this `spooky action at a distance' has made Einstein thought that quantum mechanics is not a complete \cite{ein}, the observations of quantum entanglement have been continuously demonstrated in experiments with linear photons system \cite{pan1,panjw,ten}, cavity quantum electrodynamics (QED) system \cite{haroche,haroche1}, and trapped ions systems \cite{ion1,ion2,monz}, etc. Nowadays, quantum entanglement, as a fundamental feature in quantum mechanics, has greatly promoted the development of quantum information.

There are several entangled states that appear often in theory and experiments. For two qubits, the four maximally entangled Bell states form a complete orthonormal basis of the Hilbert space \cite{Bell}, which play a fundamental role in Bell's theorem, and are also known as EPR pairs in quantum key distribution protocols \cite{long,deng}. For three qubits or more, there are two inequivalent classes of
maximally entangled states such
as Greenberger-Horne-Zeilinger (GHZ) and W states, both of them provide stronger refutations of local realism
and are more useful in quantum information processing (QIP) \cite{ghz,wstate}. Compared with the maximal entanglement, some non-maximally entangled states possess more practical capability in certain QIP tasks. For instance, the idea of decoherence-free subspaces (DFS) was brought forward to passively prevent the quantum system against a special class of decohrence \cite{DFS1,duan,DFS2}. The quantum information encoded into DFS could keep a unitary evolution of system, since they are decoupled from the environment. Due to the above properties, quantum entanglement has become the core of quantum information science, and researchers have devoted themselves to generate various of entangled states with high quality \cite{eBell,ewstate,gw,xiu}.

An intuitive and effective way for manipulation of quantum states is to design a quantum dynamic or adiabatic process that  unitarily map an initial state to the target state. Nevertheless, the inevitable interaction between quantum system and its surrounding reservoir will destroy the coherence of quantum components, thus decoherence makes
it an obstacle to preparing faithful and reliable entanglements in experiments \cite{zoller, zurek}. Fortunately, recent developments of technologies suggest that quantum feedback strategy can be taken advantage of controlling and overcoming  entanglement degradation in open quantum system.
Using approach of quantum trajectories \cite{wiseman},  the theory of quantum-limited feedback for continuously monitored systems is characterized by a deterministic Markovian master
equation, as the time delay in the feedback loop is negligible.
This method was successfully exploited to enhance the steady-state entanglement of two atoms by homodyne-mediated feedback \cite{mil,wang}, and the amount and the robustness of entanglement were substantially improved further  via quantum-jump-based  quantum feedback \cite{1,2,3,shaoxq,ben}.

In the later direct feedback schemes \cite{1,3},
 application of nonidentical feedback Hamiltonian, breaking the symmetry properties with respect to exchange of
atoms, admits a single steady-state solution of the master equation for system. As a result, a maximally entangled state is always achievable from an arbitrary initial state. However, we note that the output entangled state is closely related to the angular momentum state with $J=0$, where $J$ is the total spin of system consisting of $n$ equivalent pseudospin-1/2 particles. This situation imposes a strict restriction on the parity of particle number $n$, which is not available to prepare any other kinds of multipartite entanglement.

In this paper, we propose an efficient scheme for stabilization of quantum-feedback-based entanglement with Rydberg atoms \cite{rd1,rd2,rd3,rd4,rd5,rd6,rd7}. The advantage for adopting
 Rydberg atoms as  qubits is twofold: On the one hand, an excited atom can cause sufficiently large
energy shifts of Rydberg states in its neighboring atoms, thus the whole system is blockaded into a single excitation subspace at most. This blockade mechanism greatly reduces the dimension of investigated system and contributes to an analytical steady-state solution for the stochastic master equation. On the other hand, the  Rydberg state with a large principle quantum number is able to live for a very long time, which admirably suits for being encoded quantum information. Furthermore, the form of entangled steady states is adjustable and the feedback control is applied simply on a regular atom, irrespective of the number of particles.

The  remainder  of the paper is organized as follows. In
Sec.~II, we derive an effective Hamiltonian of the interaction between multipartite cascade-type Rydberg atoms and a damped cavity. In Sec.~III, we obtain an effective master equation describing atomic collective amplitude damping induced by a large cavity loss. In Sec.~IV, we analytically and numerically investigate the effect of quantum feedback on preparation of bipartite-, tripartite-, and multipartite entanglement, respectively. In Sec.~V, we discuss the experimental feasibility of our proposal and give a conclusion.

\section{effective physical model}
\begin{figure}
\scalebox{0.24}{\includegraphics{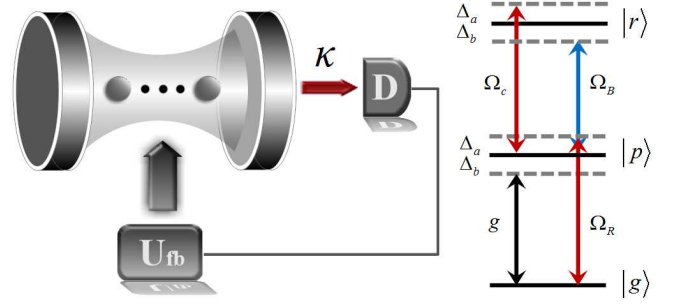} }
\caption{\label{p1}(Color online) Schematic view of the feedback setup and atomic-level
configuration. The system consists of many cascade three-level atoms simultaneously interaction with non-resonant classical fields and a quantized cavity field. Each atom is constituted by a Rydberg state $|r\rangle$, an optical state $|p\rangle$, and a ground state $|g\rangle$. The intermediate state $|p\rangle$  can be eliminated adiabatically under the large-detuning condition $|\Delta_{a(b)}|\gg \{g,\Omega_{c(B,R)}\}$, which is then reduced to an effective two-level atom resonantly coupled to a damped optical cavity with coupling strength $-g\Omega_c/\Delta_b$, and driven by a laser field with Rabi frequency $\Omega_R\Omega_B/\Delta_a$.  The single-atom feedback control $U_{\rm fb}$ is triggered right after the leakage photon is measured by the detector $D$. }
\end{figure}

We consider multipartite Rydberg atoms with cascade-type configuration
are trapped in an optical cavity, as shown in Fig.~\ref{p1}. Each atom is constituted by a Rydberg state $|r\rangle$, an optical state $|p\rangle$, and a ground state $|g\rangle$. The indirect transition from $|g\rangle$ to $|r\rangle$ mediated by $|p\rangle$ is driven by two independent channels: In one channel, the atom is first coupled to the cavity mode with strength $g$, detuned by $\Delta_b$, and then pumped by a classical field with Rabi frequency $\Omega_c$, detuning $-\Delta_a$. The other channel is totally composed by two laser fields, and the corresponding Rabi frequencies and detuings are $\Omega_R$, $-\Delta_b$, and $\Omega_B$, $\Delta_a$, respectively. All parameters are assumed to be real for the sake of simplicity.
In the interaction picture, the Hamiltonian of the system reads ($\hbar = 1$)
\begin{eqnarray}\label{full}
H_{\rm I}&=&\sum_{i=1}^N[(ge^{i\Delta_b t}a+\Omega_R^ie^{-i\Delta_a t})|p\rangle_{ii}\langle g|\nonumber\\&&+(\Omega_c^ie^{-i\Delta_b t}+\Omega_B^ie^{i\Delta_a t})|r\rangle_{ii}\langle p|
+{\rm H.c.}]\nonumber\\&&+\sum_{i<j}^NU_{ij}(r)|r\rangle_{ii}\langle r|\otimes|r\rangle_{jj}\langle r|,
\end{eqnarray}
where the Rydberg-mediated interaction $U_{ij}(r)$  arises from the dipole-dipole potential of the scale $C_3/r^3$ or the long-range van der Waals interaction proportional to $C_6/r^6$, with $r$ being the distance
between two Rydberg atoms, and $C_{3(6)}$ depending on the quantum numbers of Rydberg state. In the regime of the large detuning $|\Delta_{a(b)}|\gg \{g,\Omega_{c(B,R)}\}$, we may safely eliminate the intermediate state $|p\rangle$, and the above Hamiltonian reduces to
\begin{eqnarray}\label{a1111}
H_{\rm I}&=&\sum_{i=1}^N(\frac{\Omega_R^{i2}}{\Delta_a}-\frac{g^2}{\Delta_b}a^{\dag}a)|g\rangle_{ii}\langle g|+(\frac{\Omega_B^{i2}}{\Delta_a}-\frac{\Omega_c^{i2}}{\Delta_b})|r\rangle_{ii}\langle r|\nonumber\\
&&+[(\frac{\Omega^i_R\Omega^i_B}{\Delta_a}-\frac{g\Omega^i_c}{\Delta_b}a)|r\rangle_{ii}\langle g|+{\rm H.c.}]\nonumber\\
&&+\sum_{i<j}^NU_{ij}(r)|r\rangle_{ii}\langle r|\otimes|r\rangle_{jj}\langle r|.
\end{eqnarray}
The first two terms represent the Stark shifts of ground states and Rydberg states, respectively. Apart from canceling them via introducing other auxiliary levels, these terms can be set to commute with our prepared target steady state. Therefore, we just reserve the Raman-like transition terms and the above Hamiltonian is simplified to be
\begin{eqnarray}\label{effe}
{H}_{\rm eff}&=&\sum_{i=1}^N\Omega^i_{\rm eff}|r\rangle_{ii}\langle g|+g^i_{\rm eff}|r\rangle_{ii}\langle g|a+{\rm H.c.}\nonumber\\
&&+\sum_{i<j}^NU|r\rangle_{ii}\langle r|\otimes|r\rangle_{jj}\langle r|,
\end{eqnarray}
where $\Omega_{\rm eff}^i=\Omega_R^i\Omega_B^i/\Delta_a$, and $g_{\rm eff}^i=-g\Omega_c^i/\Delta_b$. It is reasonable to replace the distance-related Rydberg mediated interaction strength with an identical $U={\rm min}\{U_{ij}(r)\}$, because the blockade effect merely depends upon the minimum of all $U_{ij}(r)$. Eq.~(\ref{effe}) describes the interaction between multipartite effective two-level atoms and a cavity field, simultaneously driven by classical laser fields. One may also  choose a two-level configuration of the trapped atom right from the start, but the possible benefit of starting from a cascade-type atomic configuration is that the effective atom-cavity interaction is tunable via modulating the corresponding detunings and Rabi frequencies, which provides more feasibility for experimental control.

\section{dissipative dynamics of multipartite Rydberg atoms}
The dissipation
channels of the present physical model include the spontaneous emission of Rydberg state (symbolled as
$\gamma_r$)
and photon loss of the cavity mode (symbolled as $\kappa$). Under the assumptions
that the decay channels are independent, the master equation
of the whole system can be expressed by the Lindblad form \cite{scu}

\begin{eqnarray}\label{www}
\dot{\rho}&=&-i[H_{\rm eff},\rho]+\frac{\kappa}{2}(2a\rho a^{\dag}-a^{\dag}a\rho-\rho a^{\dag}a)\nonumber\\
&&+\sum_{i=1}^N\frac{\gamma_r}{2}(2\sigma^i_-\rho \sigma^i_{+}-\sigma^i_{+}\sigma^i_-\rho-\rho \sigma^i_{+}\sigma^i_-),
\end{eqnarray}
where $\sigma^i_-=\sigma^{i\dag}_+=|g\rangle_{ii}\langle r|$ is the lowering operator of a single atom. In a strong Rydberg blockade regime, the above Lindblad master equation is reduced to
\begin{eqnarray}\label{vvv}
\dot{\rho}^r&=&-i\Omega_{\rm eff}[(J_l^++J_l^-),\rho^r]-i g_{\rm eff}[(J_c^+a+J_c^-a^{\dag}),\rho^r]\nonumber\\&&+\sum_{i=1}^N\gamma{\cal D}[\sigma^i_-]\rho^r+\kappa{\cal D}[a]\rho^r,
\end{eqnarray}
where $\rho^r$ stands for the density matrix of system without considering the double occupations of Rydberg states or more. The corresponding coupling strengths are scaled by $\Omega_{\rm eff}$$=$${\rm min}\{\Omega^i_{\rm eff}\}$ and $g_{\rm eff}$$=$${\rm min}\{g^i_{\rm eff}\}$. $J_l^-=|g_1$$\dots$$g_i$$\dots$$g_N\rangle$$\sum_{i=1}^N$$\Omega^i_{\rm eff}/\Omega_{\rm eff}$$\langle g_1$$\dots r_i$$\dots$$g_N|$ and $J_c^-=|g_1$$\dots$$g_i$$\dots$$g_N$$\rangle$$\sum_{i=1}^N$$g^i_{\rm eff}/g_{\rm eff}$$\langle g_1$$\dots r_i$$\dots$$g_N|$ represent the collective lowing operators related to the classical fields and  the quantized cavity mode, respectively. $\gamma{\cal D}[\sigma^i_-]$ and $\kappa{\cal D}[a]$ denote the superoperators describing decay of atom and cavity, respectively.

In order to gain a better insight into the dissipative dynamics of system, we now rewrite the density operator in the photon number representation \cite{wang}, i.e.
\begin{eqnarray}\label{222}
\rho^r&=&\sum_{m,n=0}^\infty\rho^r_{mn}|m\rangle\langle n|,
\end{eqnarray}
where $\rho^r_{mn}$ are the density matrix elements in the basis of the photon number
states with respect to the cavity mode. For a strongly damped cavity mode,
the highly excited cavity modes only  act as perturbations, thus an
expansion to $m,n=1$ is good enough for our concerns. After substituting the above equation into $\dot{\rho}^r$,  we obtain
a set of coupled equations for the cavity field matrix
elements:
\begin{equation}\label{e1}
\dot{\rho}^r_{00}={\cal L}\rho^r_{00}-i g_{\rm eff}[J_c^+\rho^r_{10}-\rho^r_{01}J_c^-]
+\kappa\rho^r_{11},
\end{equation}
\begin{equation}\label{e2}
\dot{\rho}^r_{10}={\cal L}\rho^r_{10}-ig_{\rm eff}[J_c^-\rho^r_{00}-\rho^r_{11}J_c^-]-\frac{\kappa}{2}\rho^r_{10},
\end{equation}
\begin{equation}\label{e3}
\dot{\rho}^r_{11}={\cal L}\rho^r_{11}-i g_{\rm eff}[J_c^-\rho^r_{01}-{\rho^r}_{10}J_c^+]-\kappa\rho^r_{11},
\end{equation}
where the photon-independent terms have been absorbed into the superoperator ${\cal L}\rho^r_{ij}$. Compared with other two terms, the coherence
${\rho}^r_{10}$ changes more slowly in time, as the most populated state of cavity mode is the vacuum state. Thus it is reasonable to take $\dot{\rho}^r_{10}=0$, and we get the value of this operator as
\begin{equation}\label{8}
\rho^r_{10}=\rho^{r\dag}_{01}\approx-\frac{2ig_{\rm eff}}{\kappa}[J_c^-\rho^r_{00}-\rho^r_{11}J_c^-].
\end{equation}
Substituting the corresponding result into $\dot{\rho}^r_{00}$ and $\dot{\rho}^r_{11}$, we find
\begin{eqnarray}\label{8}
\dot{\rho}^r_{00}&=&{\cal L}\rho^r_{00}-\frac{4g_{\rm eff}^2}{\kappa}[J_c^+J_c^-\rho^r_{00}+\rho^r_{00}J_c^+J_c^-\nonumber\\&&-2J_c^+\rho^r_{11}J_c^-]+\kappa\rho^r_{11},
\end{eqnarray}
\begin{eqnarray}\label{8}
\dot{\rho}^r_{11}&=&{\cal L}\rho^r_{11}-\frac{4g_{\rm eff}^2}{\kappa}[J_c^-J_c^+\rho^r_{11}+\rho^r_{11}J_c^-J_c^+\nonumber\\&&-2J_c^-\rho^r_{00}J_c^+]-\kappa\rho^r_{11}.
\end{eqnarray}
These two terms characterize the dynamical evolution of atoms because of ${\rho}^r_{00}+{\rho}^r_{11}={\rho}^r_{atom}$. Now we add them together and adiabatically eliminate
the element $\rho^r_{11}$, the
master equation for the reduced density operator of atoms becomes
\begin{equation}\label{bbb}
\dot{\rho}^r=-i\Omega_{\rm eff}[(J_l^++J_l^-),\rho^r]+\Gamma{\cal D}[J_c^-]\rho^r+\sum_{i=1}^N\gamma{\cal D}[\sigma^i_-]\rho^r,
\end{equation}
where $\Gamma=4g_{\rm eff}^2/\kappa$ is the collective amplitude
damping rate of the transition from $|r\rangle$ to $|g\rangle$. Supposing that the collective
decay rate is much larger than the spontaneous emission
rate $\Gamma\gg\gamma$, we may have an effective master equation of multipartite Rydberg atoms as
\begin{equation}\label{9999}
\dot{\rho}^r=-i\Omega_{\rm eff}[(J_l^++J_l^-),\rho^r]+\Gamma{\cal D}[J_c^-]\rho^r.
\end{equation}
In general, the dynamical steady solution of Eq.~(\ref{9999}) is a mixed state.
A sufficient condition that a steady state should satisfy is $\Omega_{\rm eff}^i/\Omega_{\rm eff}=g_{\rm eff}^i/g_{\rm eff}$, i.e. it is an eigenstate of the
collective quantum jump operator $J^-=J^-_c=J^-_l$ corresponding to zero eigenvalue ($[J^+,\rho^r]=0$ due to the strong Rydberg blockade effect). In the following, we will pick up a maximally steady entanglement with the help of quantum feedback control.

\section{quantum-jump-based feedback}
The quantum feedback theory is a combination of quantum measurement and master equation. For the purpose of understanding quantum feedback dynamics concisely, it is instrumental to decompose the superoperator ${\cal D}[J^-]\rho^r$ into two parts \cite{wiseman}, one part
\begin{equation}\label{999}
{\cal A}[J^-]\rho^r=\frac{1}{2}[J^+J^-\rho^r+\rho^r J^+J^-]
\end{equation}
indicats a null measurement, leaving the density operator ${\tilde{\rho}_0^r(t+dt)}$  unchanged, and the other part
\begin{equation}\label{999}
{\cal J}[J^-]\rho^r=J^-\rho^rJ^+
\end{equation}
corresponds to  a detection of signal, which is immediately followed by the feedback control in the form of
\begin{equation}\label{999}
{\tilde{\rho}_1^r(t+dt)}=e^{\cal K}J^-\rho^r(t)J^+dt,
\end{equation}
where ${\cal K}$ is a Liouville superoperator.
Since the nonselective evolution of the system is given by
\begin{equation}\label{999}
{\rho^r(t+dt)}={\tilde{\rho}_1^r(t+dt)}+{\tilde{\rho}_0^r(t+dt)},
\end{equation}
we can directly incorporate the feedback operator into the master equation of system as
\begin{equation}\label{999}
\dot{\rho}^r=-i\Omega_{\rm eff}[(J^++J^-),\rho^r]+e^{\cal K}{\cal J}[J^-]\rho^r-{\cal A}[J^-]\rho^r.
\end{equation}
In the case that
${\cal K}{\rho^r}=-i[z,\rho^r]$, we have
\begin{equation}\label{eff}
\dot{\rho}^r=-i\Omega_{\rm eff}[(J^++J^-),\rho^r]
+\Gamma{\cal D}[U_{\rm fb}J^-]\rho^r,
\end{equation}
where $U_{\rm fb}=e^{-iz}$ is the unitary operator of feedback control operating on system under the circumstance of Rydberg blockade. Note that the feedback does not alter the steady pure state solution of Eq.~(\ref{9999}), because of ${\cal D}[U_{\rm fb}J^-]\rho^r=U_{\rm fb}J^-\rho^rJ^+U_{\rm fb}^{\dag}-(J^+J^-\rho^r+\rho^rJ^+J^-)/2$. Therefore, a choice of $z$ is key to generate bipartite-, tripartite-, and multipartite entanglement.
\subsection{Bipartite entanglement}
Although the entangled states of two particles are the simplest example of entanglement, which have been realized with different approaches, we will show the specific function of quantum feedback in the present scheme, i.e. dissipative preparation of entanglement for Rydberg atoms. The previous works suggest that the principle of
selecting feedback control is to violate the symmetry with
respect to exchange of atoms, so we take the feedback operator as
\begin{equation}\label{a}
U_{\rm fb}=\exp\{-i\lambda[(|g\rangle_{11}\langle r|+|r\rangle_{11}\langle g|)\otimes I_2+\frac{U}{\lambda}|rr\rangle\langle rr|]\delta t\},
\end{equation}
where $\lambda$ denotes the feedback strength operated on the first atom, $I_2$ is the identity operator of the second atom, and we have assumed that the Rydberg blockade still works during the feedback operation.  This unitary operator can be approximated to
\begin{equation}\label{b}
U^{r}_{\rm fb}=\exp[-i\omega(|g\rangle_{11}\langle r|+|r\rangle_{11}\langle g|)\otimes|g\rangle_{22}\langle g|],
\end{equation}
which represents the finite amount of evolution $(\omega=\lambda \delta t)$
imposed by the control Hamiltonian on the system under the condition of $U\gg\lambda$. Thus a single-qubit flip operation on the first atom is equivalent to a controlled-flip operation in the strong Regime of Rydberg blockade. In order to find the stationary solution of Eq.~(\ref{eff}), we are encouraged to expand the density operator $\rho^r$ in a subspace spanned by
\begin{eqnarray}\label{11}
|1\rangle_{\rm B}&=&|gg\rangle,\nonumber\\|2\rangle_{\rm B}&=&\frac{1}{\sqrt{2}}(|gr\rangle+|rg\rangle),\nonumber\\
|3\rangle_{\rm B}&=&\frac{1}{\sqrt{2}}(|gr\rangle-|rg\rangle).
\end{eqnarray}
After setting $\Omega_{\rm eff}=\Omega^i_{\rm eff}$ and $\dot{\rho}^r=0$ in Eq.~(\ref{eff}), we acquire an analytical expression for the corresponding algebraic equation
\begin{figure*}
\begin{minipage}[t]{0.42\linewidth}
\centering
\includegraphics[width=2.6in]{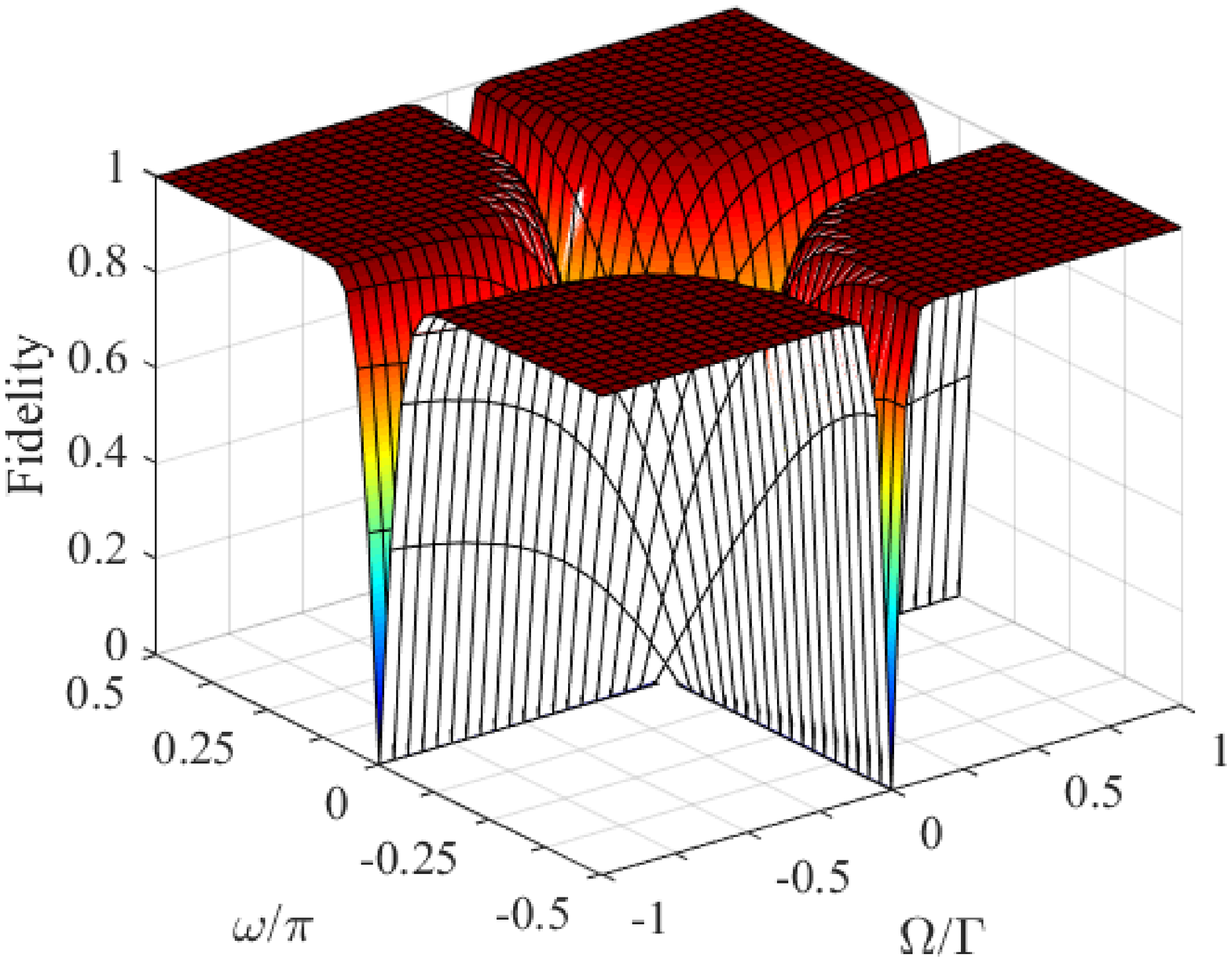}
 \centerline{(a)}
\end{minipage}%
\begin{minipage}[t]{0.42\linewidth}
\centering
\includegraphics[width=2.4in]{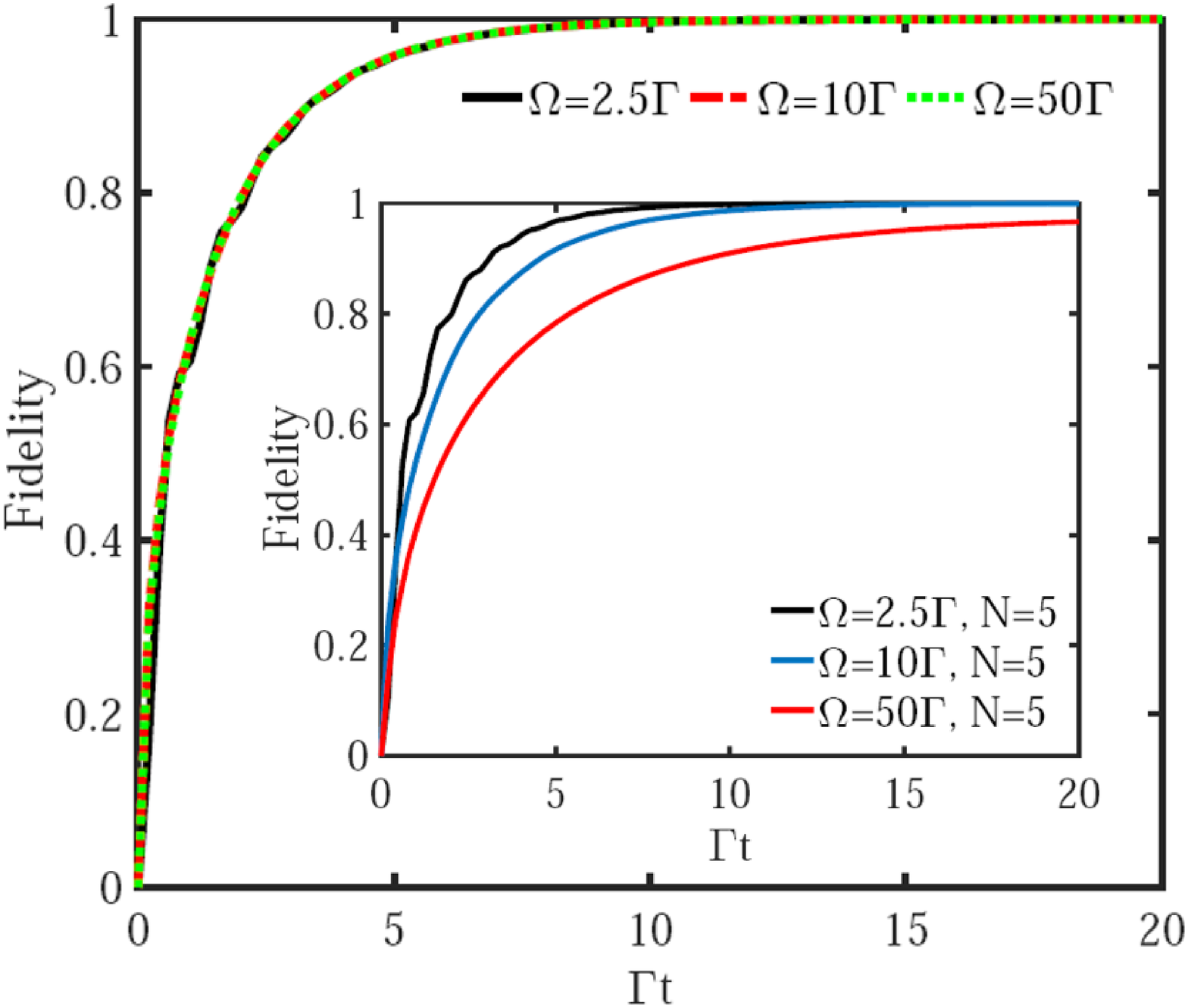}
 \centerline{(b)}
\end{minipage}
\begin{minipage}[t]{0.42\linewidth}
\centering
\includegraphics[width=2.4in]{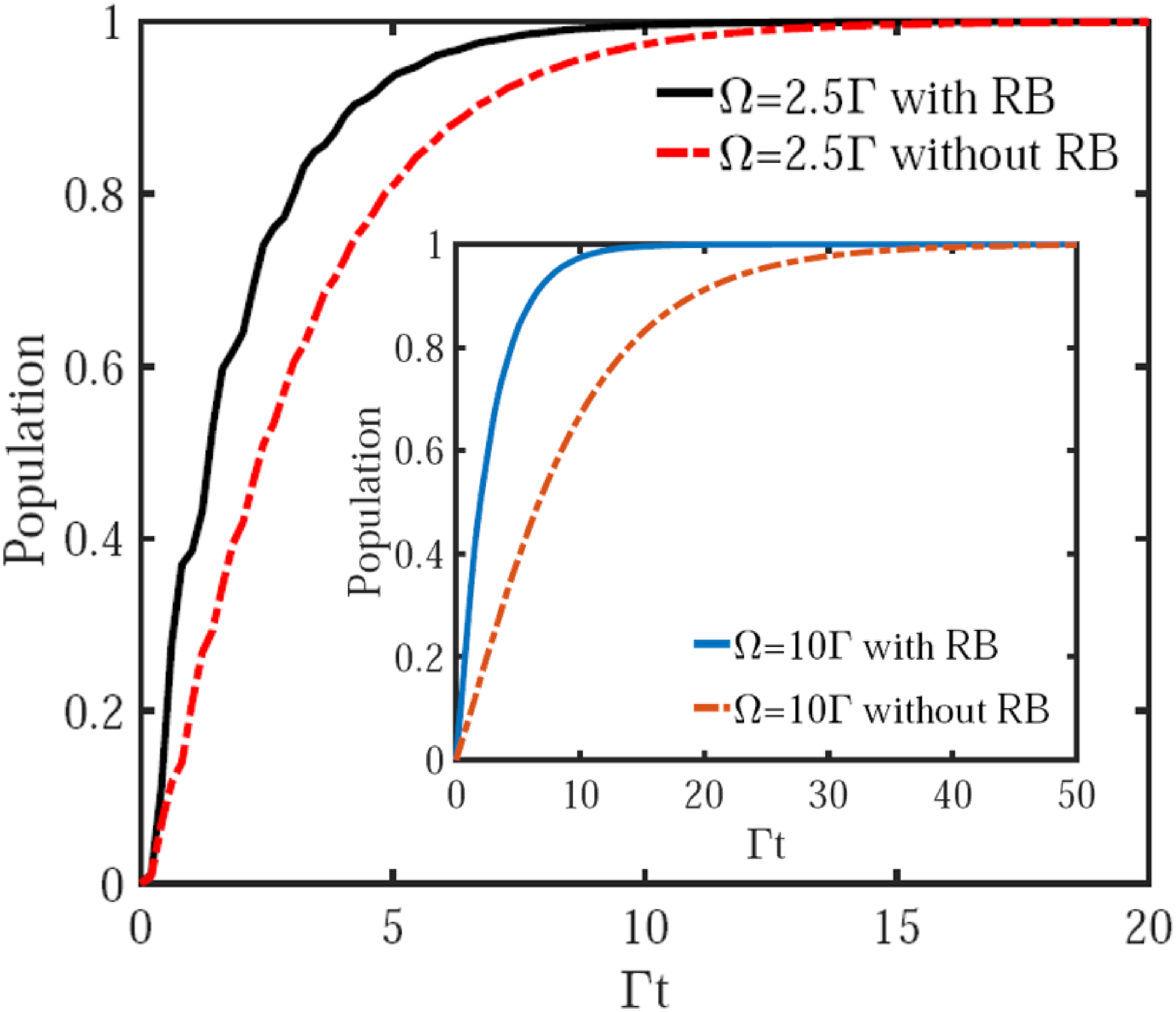}
 \centerline{(c)}
\end{minipage}%
\begin{minipage}[t]{0.42\linewidth}
\centering
\includegraphics[width=2.6in]{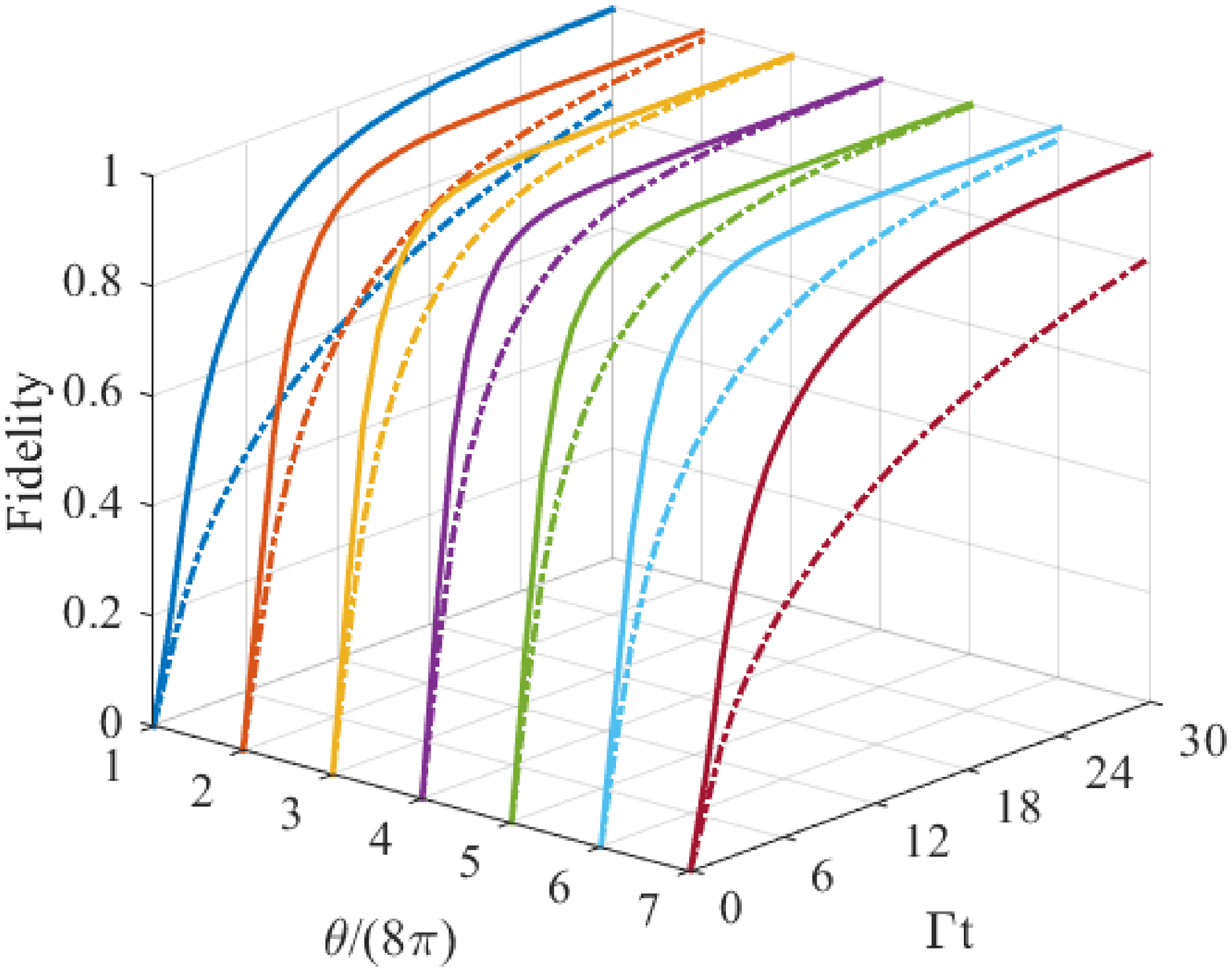}
 \centerline{(d)}
\end{minipage}
\caption{\label{p2}(Color online) Analysis of bipartite entanglement under quantum feedback conrol. (a) The target-state fidelity obtained from the effective master equation of Eq.~(\ref{eff}) (undersurface), which is in excellent agreement with result derived from the original master equation of Eq.~(\ref{ful}) (upper surface), under the given parameters $\kappa=25\Gamma$, $g_{\rm eff}=2.5\Gamma$, $U=500\Gamma$, $t=100/\Gamma$. (b) The fidelity calculated from the effective master equation is not consistent with the original one (inset), out of weak driving regime, ($N$ denotes the considered dimension of cavity mode) where we have chosen $\omega=0.5\pi$, and other parameters are the same as Fig.~\ref{p2}(a). (c) The Rydberg blockade provides a way to speed up convergence of target-state population originating from a two-atom ground state $|gg\rangle$ (solid line), compared with the population in the absence of Rydberg blockade (dotted-dashed line). This speedup effect will be more pronounced for a stronger driving field $\Omega=10\Gamma$ as shown in the inset. (d) Time evolution of the fidelity for the target-state $(\cos\theta|gr\rangle+\sin\theta|rg\rangle)$ with unity (solid line) and 0.5 (dotted-dashed line) detection efficiency, respectively. The effective Rabi frequency has been set as $\Omega=0.25\Gamma$.}
\end{figure*}
\begin{widetext}
\begin{align}\label{bb}
\begin{bmatrix}\begin{smallmatrix}
-i\sqrt{2}(\rho^r_{12}-\rho^r_{21})\Omega-2\rho^r_{22}\Gamma\cos\omega^2 &
 \rho^r_{12}\Gamma-i\sqrt{2}\big[(\rho^r_{11}-\rho^r_{22})\Omega+\rho^r_{22}\Gamma\cos\omega\sin\omega\big] & i\sqrt{2}(\rho^r_{23}\Omega+\rho^r_{22}\Gamma\cos\omega\sin\omega) \\
 \rho^r_{21}\Gamma+i\sqrt{2}\big[(\rho^r_{11}-\rho^r_{22})\Omega+\rho^r_{22}\Gamma\cos\omega\sin\omega\big]&
i\sqrt{2}(\rho^r_{12}-\rho^r_{21})\Omega-\rho^r_{22}\Gamma(\sin\omega^2-2) &
i\sqrt{2}\rho^r_{13}\Omega+\Gamma(\rho^r_{23}+\rho^r_{22}\sin\omega^2)\\
 -i\sqrt{2}(\rho^r_{32}\Omega+\rho^r_{22}\Gamma\cos\omega\sin\omega) &
  -i\sqrt{2}\rho^r_{31}\Omega+\Gamma(\rho^r_{32}+\rho^r_{22}\sin\omega^2)&
  -\rho^r_{22}\Gamma\sin\omega^2\\
\end{smallmatrix}\end{bmatrix}=0,
\end{align}
\end{widetext}
where the subscript of Rabi frequency is omitted.

 Let us now analyze the nontrivial solution of Eq.~(\ref{bb}). The feedback strength $\omega$ is an adjustable parameter. If we restrict $\sin\omega\neq0$, a straightforward calculation shows that all elements in Eq.~(\ref{bb}) are zeros. Keeping in mind the feedback operator does not violate the conservation of probability, i.e. $\rho^r_{11}+\rho^r_{22}+\rho^r_{33}=1$, so we can conclude that the antisymmetric Bell state $|3\rangle=(|gr\rangle-|rg\rangle)/\sqrt{2}$ is the unique steady solution. In experiments, the signal does not come from the collective damping of atoms, but monitoring the photon leakage out of the cavity mode. Thus the correctness of adiabatic approximation made in Eq.~(\ref{eff}) can be determined by considering a more realistic feedback master equation
\begin{eqnarray}\label{ful}
\dot{\rho}={\cal L }\rho-i g_{\rm eff}[(J^{+}a+J^-a^{\dag}),\rho]+\kappa{\cal D}[U_{\rm fb}a]\rho,
\end{eqnarray}
where ${\cal L}\rho=-i\Omega[(J^++J^-),\rho]-iU[|rr\rangle\langle rr|,\rho]$. This model is able to be transformed back to Eq.~(\ref{eff}) if an adiabatic
elimination is performed. In order to measure the distance between quantum states, we adopt the definition of fidelity $F(\sigma,\rho(t))\equiv{\rm Tr}\sqrt{\sigma^{1/2}\rho(t)\sigma^{1/2}}$ with $\sigma$ being the target state \cite{Nielsen}. In Fig.~\ref{p2}(a), we initialize the system into state $|gg\rangle$ and plot the target-state fidelity as a function of feedback strength $\omega$
and effective Rabi frequency $\Omega$ with the effective master equation of Eq.~(\ref{eff}) (undersurface), and the original master equation of Eq.~(\ref{ful}) (upper surface). These two results are in excellent agreement with each other under the given parameters $\kappa=25\Gamma$, $g_{\rm eff}=2.5\Gamma$, $U=500\Gamma$, $t=100/\Gamma$, which in turn proves the rationality of the above adiabatic approximation. It can also be seen that the present scheme is robust again the fluctuation of parameters, as the fidelity maintains unity for a wide range of $\omega$ and $\Omega$.

In Fig.~\ref{p2}(b),
we investigate the effect of strong driving on the convergent time for fidelity with Eqs.~(\ref{eff}) and (\ref{ful}), respectively. For the effective model, there exists a limit value for $\Omega$, beyond which the asymptotic time is almost unchanged. But this is not the case in reality, as shown in the inset of Fig~\ref{p2}(b), where we have considered four-photon excitation in the cavity mode ($N=5$) to provide a more distinct physical picture. In fact,
the stationary entanglement is produced under the competitions of classical driving, quantum feedback and dissipation. The numerical simulation illustrates that an better value (guaranteeing a shorter time for reaching the target state) of Rabi frequency for driving field should be modulated to the same order of magnitude of $g_{\rm eff}$. Thus our results reveal that in the presence of strong driving fields, the effective master equation of Eq.~(\ref{eff}) is not suitable for describing dynamical evolution of system, but the feedback mechanism remains established from the viewpoint of steady state. In what follows, our simulations are all based on the original master equation Eq.~(\ref{ful}) without any specification.

Fig.~\ref{p2}(c) displays one of the superiority of Rydberg atoms for implementing quantum-feedback-based entanglement. The solid line and the dashed-dotted line represent the populations of target states with and without considering Rydberg Blockade, respectively. For weak driving fields $\Omega=2.5\Gamma=g_{\rm eff}$, the solid line exceeds $99\%$ at a short time $\Gamma t=9$, while the dashed-dotted line is just about $96\%$. The gap between the above populations is further broadened in the regime of strong driving fields ($\Omega=10\Gamma$), as shown from the inset. For $\Gamma t=16$, the solid line is almost stabilized to $99.68\%$, but the dashed-dotted line merely rises to $84.91\%$. The physical principle behind this phenomenon is that the strong Rydberg blockade excludes a simultaneous population of double Rydberg states, resulting in closure of a transition channel. Thence the concerning rate of stabilization is accelerated from an initial state to the target steady state.

Another superiority of our scheme is shown in Fig.~\ref{p2}(d). We are able to engineer an arbitrary state $(\cos\theta|gr\rangle+\sin\theta|rg\rangle)$ as the stationary state via modulating the real Rabi frequencies of classical fields. Comparatively speaking, the previous proposals are concerned only with the maximal entanglement. The preparation of superposition between states $|gr\rangle$ and $|rg\rangle$ is especially important for quantum coding, since the logic qubit $|0\rangle_L\equiv|gr\rangle$ and $|1\rangle_L\equiv|rg\rangle$ are robust against the phase error caused by $H_{de}=\epsilon_g(\tau)|g\rangle\langle g|+\epsilon_r(\tau)|r\rangle\langle r|$. For simplicity, we opt several values of $\theta$ and plot the corresponding evolutions of fidelities. (a relative phase between $|gr\rangle$ and $|rg\rangle$ is also realizable by simply introducing some complex Rabi frequencies). The dashed-dotted lines of Fig.~\ref{p2}(d) correspond to the case of imperfect detections, which are governed by
\begin{eqnarray}\label{HHH}
\dot{\rho}&=&{\cal L }\rho-i g_{\rm eff}[(J^{+}a+J^-a^{\dag}),\rho]+\eta\kappa{\cal D}[e^{-iz}a]\rho\nonumber\\
&&+(1-\eta)\kappa{\cal D}[a]\rho,
\end{eqnarray}
where $\eta$ represents the efficiency of the detector and $(1-\eta)$ indicates the case no feedback control is activated. These results show that the efficiency of the detector  will delay the convergent time of entanglement, but unaffect the quality of final state.
\subsection{Tripartite entanglement}
In this part, we mainly discuss the possibility of generating tripartite entanglement for Rydberg atoms under quantum feedback control. Referring to the case of bipartite entanglement, we first introduce four quantum states below as a new set of density operator up to single excitation,
\begin{figure}
\includegraphics[width=3.4in]{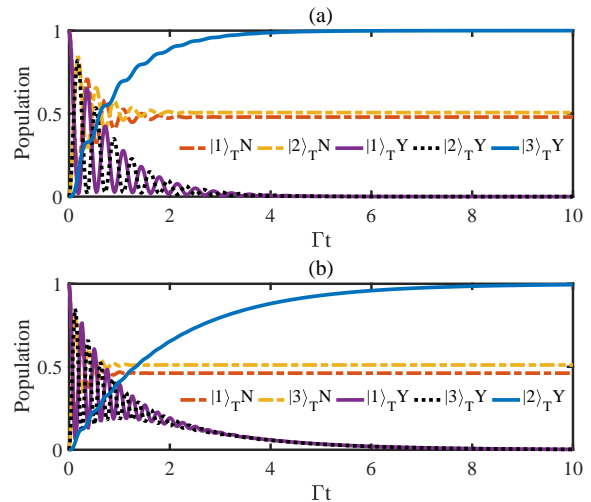}
\caption{\label{p3}(Color online) The populations of quantum states are simulated with and without quantum feedback control during the preparation of three-qubit DFS state (a) and W state (b), where $\omega=0.5\pi$, $\kappa=100\Gamma$, $g_{\rm eff}=5\Gamma$, $\Omega=5\Gamma$, and $U=2500\Gamma$.}
\end{figure}
\begin{eqnarray}\label{11}
|1\rangle_{\rm T}&=&|ggg\rangle,\nonumber\\|2\rangle_{\rm T}&=&\frac{1}{\sqrt{3}}(|ggr\rangle+|grg\rangle+|rgg\rangle),\nonumber\\
|3\rangle_{\rm T}&=&\frac{1}{\sqrt{6}}(|ggr\rangle+|grg\rangle-2|rgg\rangle),\nonumber\\
|4\rangle_{\rm T}&=&\frac{1}{\sqrt{2}}(|ggr\rangle-|grg\rangle),
\end{eqnarray}
where $|2\rangle_{\rm T}$, $|3\rangle_{\rm T}$, and $|4\rangle_{\rm T}$ constitute a complete basis for the single excitation subspace. States $|2\rangle_{\rm T}$ and $|3\rangle_{\rm T}$ are the three-qubit W state and the three-qubit DFS state against collective amplitude damping, both of them belong to the tripartite entanglement of interest.  But the state $|4\rangle_{\rm T}$ will
not participate in the process of  evolution and it can be excluded further. This point can be seen more clearly with the dissipative model of Eq.~(\ref{9999}). In the absence of quantum feedback, a quantum state initialized in $|1\rangle_{\rm T}$ can only be pumped to the collective single excitation state $|2\rangle_{\rm T}$ which then decays back to the ground state $|1\rangle_{\rm T}$, and this process repeats again and again until a dynamic equilibrium is achieved. As shown in Fig.~{\ref{p3}}(a), the upper (lower) dashed-dotted line represents the stationary population of $|2\rangle_{\rm T}$ ($|1\rangle_{\rm T}$) without applying quantum feedback. Even in the presence of quantum feedback, the operation on the first atom will not change the relative phase between other two atoms. Thus state $|4\rangle_{\rm T}$ is safely disregarded, and the density operator of system is expanded with other three states, whose steady state is solved as
\begin{widetext}
\begin{align}\label{tt}
\begin{bmatrix}\begin{smallmatrix}
-i\sqrt{3}(\rho^r_{12}-\rho^r_{21})\Omega-3\rho^r_{22}\Gamma\cos\omega^2 &
 \frac{3}{2}\rho^r_{12}\Gamma-i\sqrt{3}\big[(\rho^r_{11}-\rho^r_{22})\Omega+\rho^r_{22}\Gamma\cos\omega\sin\omega\big] & i\sqrt{3}(\rho^r_{23}\Omega+\sqrt{2}\rho^r_{22}\Gamma\cos\omega\sin\omega) \\
 \frac{3}{2}\rho^r_{21}\Gamma+i\sqrt{3}\big[(\rho^r_{11}-\rho^r_{22})\Omega+\rho^r_{22}\Gamma\cos\omega\sin\omega\big]&
i\sqrt{3}(\rho^r_{12}-\rho^r_{21})\Omega-\rho^r_{22}\Gamma(\sin\omega^2-3) &
i\sqrt{3}\rho^r_{13}\Omega+\Gamma(\frac{3}{2}\rho^r_{23}+\sqrt{2}\rho^r_{22}\sin\omega^2)\\
 -i\sqrt{3}(\rho^r_{32}\Omega+\sqrt{2}\rho^r_{22}\Gamma\cos\omega\sin\omega) &
  -i\sqrt{3}\rho^r_{31}\Omega+\Gamma(\frac{3}{2}\rho^r_{32}+\sqrt{2}\rho^r_{22}\sin\omega^2)&
  -2\rho^r_{22}\Gamma\sin\omega^2\\
\end{smallmatrix}\end{bmatrix}=0.
\end{align}
\end{widetext}
 Through the same analysis of the process as previously on bipartite entanglement, we find the three-qubit DFS state $|3\rangle_{\rm T}$ is the unique solution of Eq.~(\ref{tt}) for a nonzero $\sin\omega$, of which the dynamic evolution is characterized by the uppermost solid line of Fig.~\ref{p3}(a). Interestingly, if we modulated the Rabi frequencies of classical fields so satisfy $-\Omega_{\rm eff}^1=2\Omega_{\rm eff}^2=2\Omega_{\rm eff}^3=2\Omega_{\rm eff}$, and $-g_{\rm eff}^1=2g_{\rm eff}^2=2g_{\rm eff}^3=2g_{\rm eff}$, the roles of states $|2\rangle_{\rm T}$ and $|3\rangle_{\rm T}$ is interchanged, and the tripartite W state becomes the stationary state of system, illustrated by the uppermost solid line of Fig.~\ref{p3}(b).

\subsection{Multipartite entanglement}
The generalization from the tripartite entanglement to the multipartite entanglement is straightforward. The corresponding closed subspace with identical $\Omega$ and $g_{\rm eff}$  is spanned by
\begin{figure}
\includegraphics[width=3.2in]{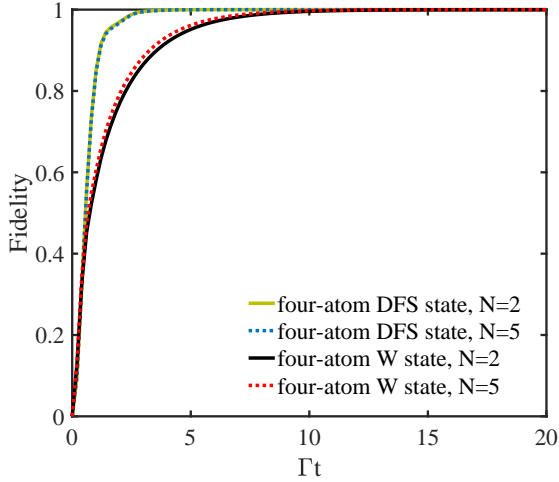}
\caption{\label{p4}(Color online) The fidelities of four-qubit DFS state and W state are simulated numerically with dimension of cavity mode $N=2$ (solid line) and $N=5$ (dotted line), where $\omega=0.5\pi$, $\Omega=\Gamma$, $\kappa=25\Gamma$, $g_{\rm eff}=2.5\Gamma$, $U=500\Gamma$.}
\end{figure}
\begin{eqnarray}\label{11}
|1\rangle_{\rm M}&=&|gg\dots g\rangle,\nonumber\\|2\rangle_{\rm M}&=&\frac{1}{\sqrt{n}}(|gg\dots r\rangle+|g\dots r\dots g\rangle+|rg\dots g\rangle),\nonumber\\
|3\rangle_{\rm M}&=&\frac{1}{\sqrt{n(n-1)}}(|gg\dots r\rangle+|g\dots r\dots g\rangle\nonumber\\&&-(n-1)|r\dots gg\rangle).
\end{eqnarray}
Besides state $|3\rangle_{\rm M}$,  there are additional $(n-2)$ degenerate quantum states, under the action of $J^-$, in the single excitation subspace. But these quantum states contribute noting for our system. So the steady-state solution of the multipartite feedback master equation is given by
\begin{widetext}
\begin{align}\label{mm}
\begin{bmatrix}\begin{smallmatrix}
-i\sqrt{n}(\rho^r_{12}-\rho^r_{21})\Omega-n\rho^r_{22}\Gamma\cos\omega^2 &
 \frac{n}{2}\rho^r_{12}\Gamma-i\sqrt{n}\big[(\rho^r_{11}-\rho^r_{22})\Omega+\rho^r_{22}\Gamma\cos\omega\sin\omega\big] & i\sqrt{n}(\rho^r_{23}\Omega+\sqrt{n-1}\rho^r_{22}\Gamma\cos\omega\sin\omega) \\
 \frac{n}{2}\rho^r_{21}\Gamma+i\sqrt{n}\big[(\rho^r_{1,1}-\rho^r_{22})\Omega+\rho^r_{22}\Gamma\cos\omega\sin\omega\big]&
i\sqrt{n}(\rho^r_{12}-\rho^r_{21})\Omega-\rho^r_{22}\Gamma(\sin\omega^2-n) &
i\sqrt{n}\rho^r_{13}\Omega+\Gamma(\frac{n}{2}\rho^r_{23}+\sqrt{n-1}\rho^r_{22}\sin\omega^2)\\
 -i\sqrt{n}(\rho^r_{32}\Omega+\sqrt{n-1}\rho^r_{22}\Gamma\cos\omega\sin\omega) &
  -i\sqrt{n}\rho^r_{31}\Omega+\Gamma(\frac{n}{2}\rho^r_{32}+\sqrt{n-1}\rho^r_{22}\sin\omega^2)&
  -(n-1)\rho^r_{22}\Gamma\sin\omega^2\\
\end{smallmatrix}\end{bmatrix}=0.
\end{align}
\end{widetext}
This equation signifies that the multipartite DFS state $|3\rangle_{\rm M}$ is the unique steady state of system in the presence of $\sin\omega\neq0$, and this steady state is able to be transformed into the multipartite W state via adjustment of Rabi frequencies of classical driving fields. As an example, we plot the fidelities of four-qubit DFS state (upper lines) and W state (lower lines) in Fig.~\ref{p4}. Under the parameters $\Omega=\Gamma$, $\kappa=25\Gamma$, $g_{\rm eff}=2.5\Gamma$, $U=500\Gamma$, it is precise enough for cutting off the dimension of cavity mode to $N=2$ (solid line), since the results agree well with the case of $N=5$ (dotted line). Eq.~(\ref{mm}) is the key finding of our work, because it provides an analytical expression for stabilization of multipartite entanglement for  Rydberg atom with quantum feedback control. By setting $n=2$ or $n=3$, the expression of Eqs~(\ref{bb}) or (\ref{tt}) is recovered.
\begin{figure*}
\begin{minipage}[t]{0.42\linewidth}
\centering
\includegraphics[width=2.6in]{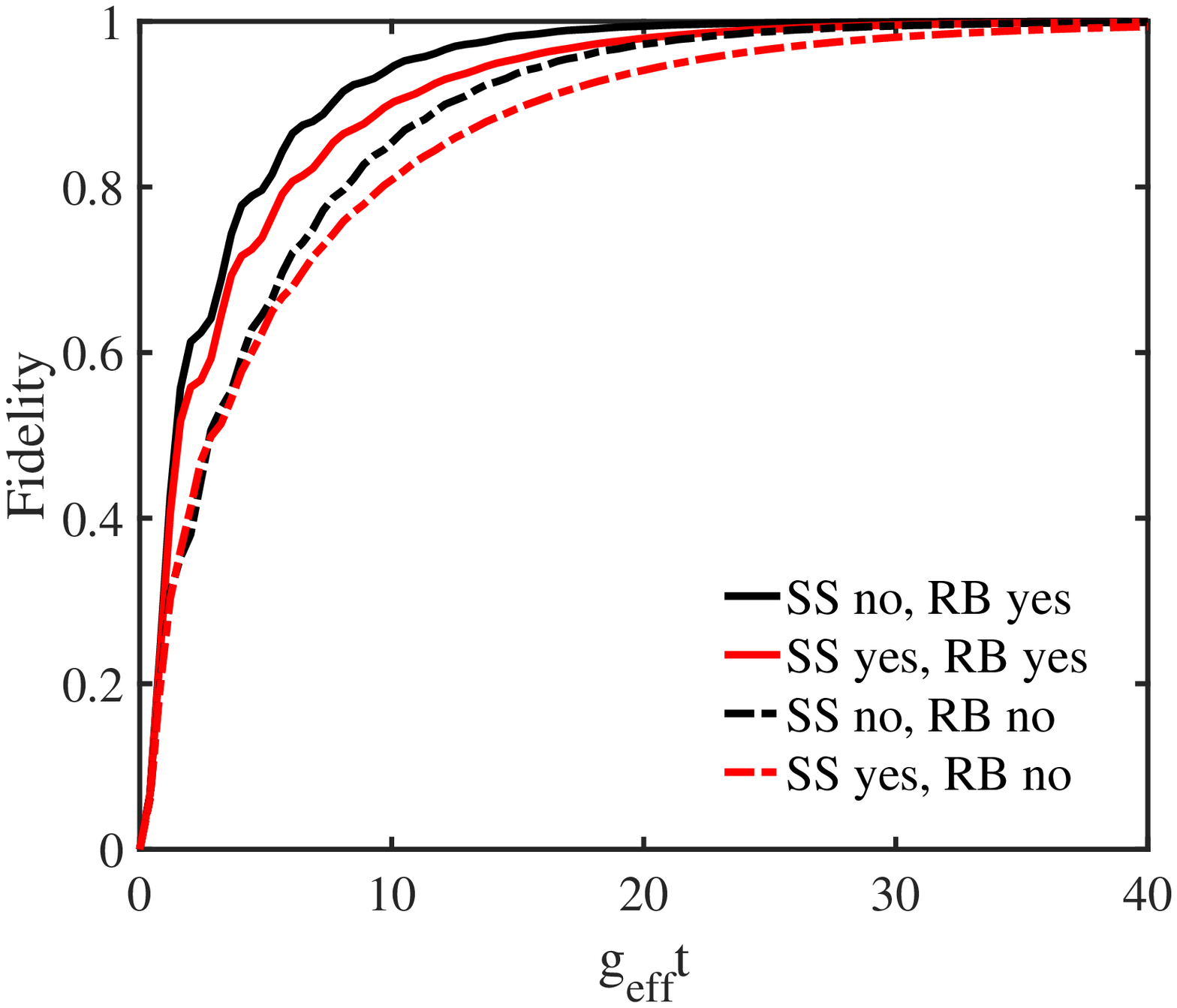}
 \centerline{(a)}
\end{minipage}%
\begin{minipage}[t]{0.42\linewidth}
\centering
\includegraphics[width=2.6in]{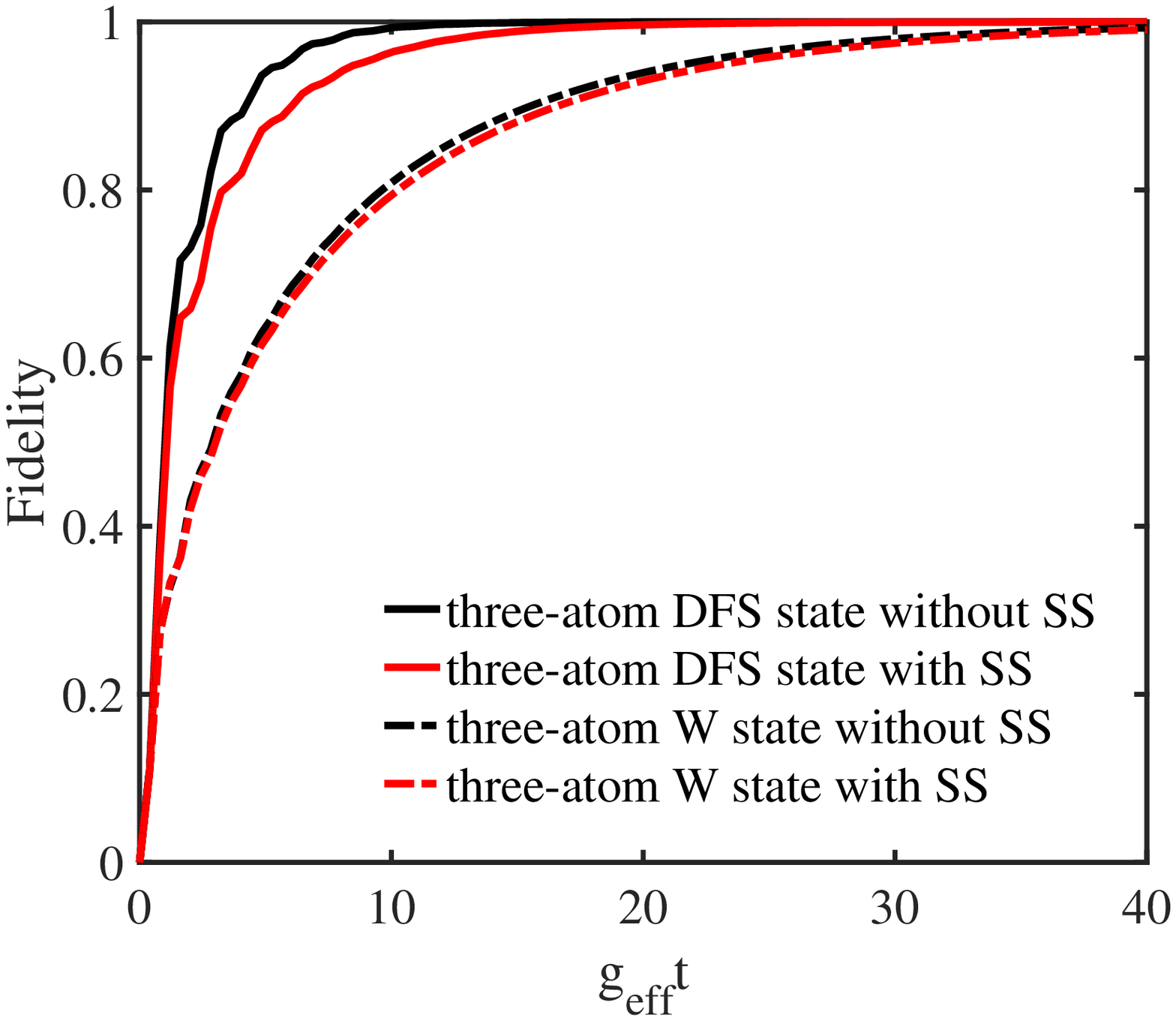}
 \centerline{(b)}
\end{minipage}
\begin{minipage}[t]{0.42\linewidth}
\centering
\includegraphics[width=2.38in]{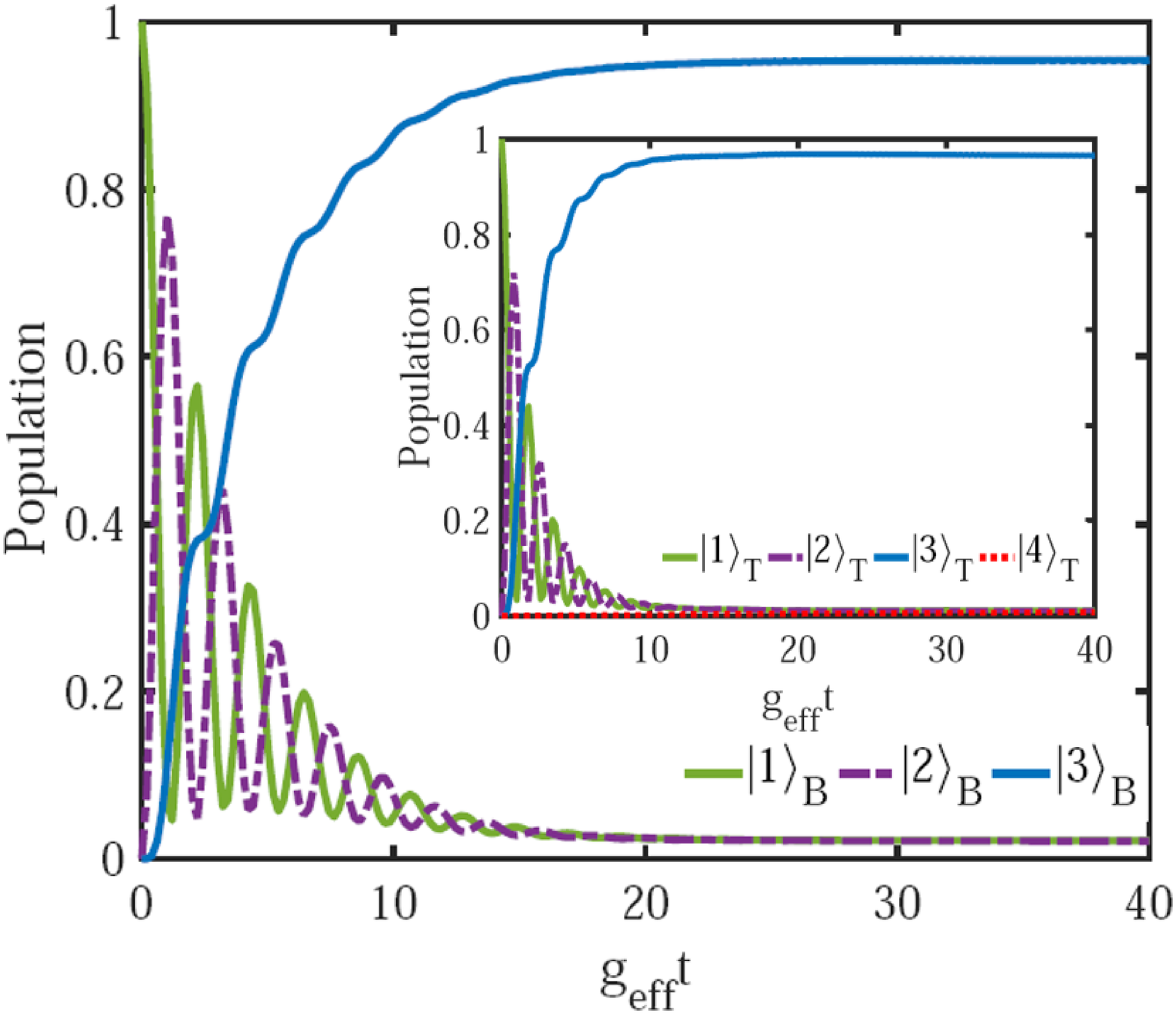}
 \centerline{(c)}
\end{minipage}%
\begin{minipage}[t]{0.44\linewidth}
\centering
\includegraphics[width=2.6in]{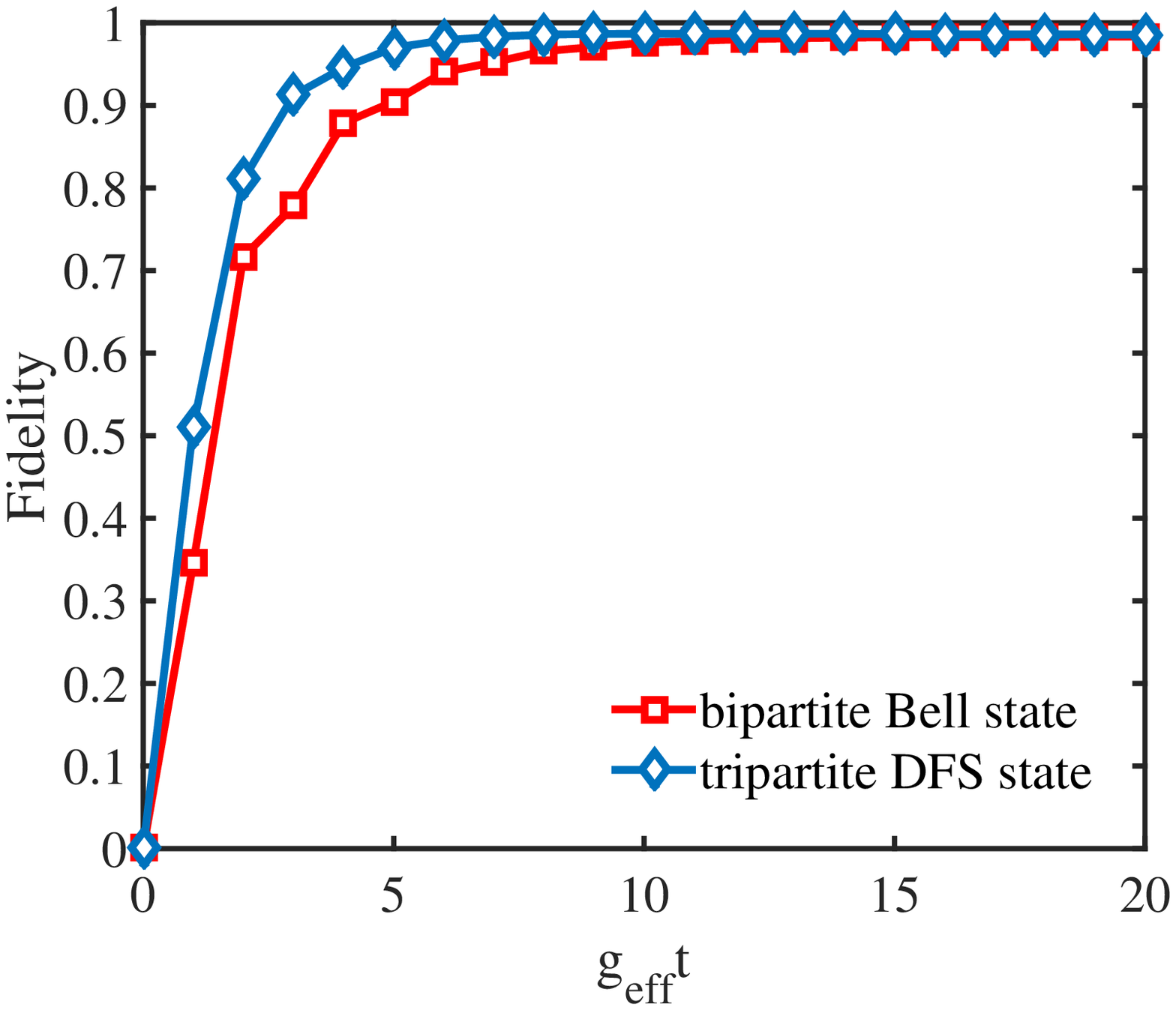}
 \centerline{(d)}
\end{minipage}
\caption{\label{p5}(Color online) (a) The effects of Stark shifts and Rydberg blockade on preparation of antisymmetric Bell state $(|gr\rangle-|rg\rangle)/\sqrt{2}$, where $\omega=0.5\pi$, $\Omega=g_{\rm eff}$, $\kappa=10g_{\rm eff}$, and $U=200g_{\rm eff}$.
(b) The influences of Stark shifts on preparation of tripartite DFS state (solid line) and tripartite W states (dashed-dotted line), with the same parameters to Fig.~\ref{p5}(a). (c) Populations of quantum states versus time during generation of the Bell state (tripartite DFS state) with spontaneous emission rates $\gamma_r=0.008g_{\rm eff}$, and $\gamma_p=8g_{\rm eff}$. (d) Fidelities of bipartite Bell state and tripartite DFS state versus time under experimental parameters.}
\end{figure*}
\section{discussions and conclusions}
The physical model mentioned above has gotten rid of the stark-shift terms, as they have no effect on current scheme from the perspective of the steady-state solution. Without loss of generality, we will take the case of realizing tripartite entanglement to support the statement. As shown in Eq.~(\ref{a1111}), the effective Rabi frequency and atom-cavity interaction are determined by $\Omega_{\rm eff}^i=\Omega_R^i\Omega_B^i/\Delta_a$ and $g_{\rm eff}^i=-g\Omega_c^i/\Delta_b$. The generation of three-qubit DFS state $|3\rangle_{\rm T}$ require the condition  $\Omega_{\rm eff}^i=\Omega_{\rm eff}$ and $g_{\rm eff}^i=g_{\rm eff}$. Although there are more than one way for selecting parameters to accomplish this goal, we choose $\Omega_{\rm B}^i=\Omega_{\rm B}$, $\Omega_{\rm R}^i=\Omega_{\rm R}$, $\Omega_{\rm c}^i=\Omega_{\rm c}$, $|\Omega_{\rm B}^i|=|\Omega_{\rm c}^i|$ and $\Delta_a=\Delta_b$. Note that the last two terms are crucial for our proposal as they automatically counteract the Stark shifts of Rydberg states $|r\rangle$. Now all atoms see the same Stark shift of ground state $|g\rangle$, leading to the relation $[|3\rangle_{\rm T}\langle 3|, (\Omega_{\rm R}^2-g^2a^{\dag}a)/\Delta]=0$. Thus the three-qubit DFS state $|3\rangle_{\rm T}$ remains the steady state of system in the presence of Stark shifts. Nevertheless, we find that the Stark shifts does affect the dynamic evolution of quantum states. Fig.~\ref{p5}(a) demonstrates the effects of Stark shifts and Rydberg blockade on preparation of bipartite entanglement $(|gr\rangle-|rg\rangle)/\sqrt{2}$. Compared with other three cases, the fidelity with Rydberg blockade but without Stark shift converges to unity fastest (upper solid line). The influences of Stark shifts on the preparation of tripartite DFS state (solid line) and tripartite W states (dashed-dotted line) are also displayed in Fig~\ref{p5}(b), which still play the role of retarding convergence.

Next we study the effect of spontaneous emission on the current proposal. In order to fully characterize the dissipative factors, we must introduce the following time-dependent master equation
\begin{eqnarray}\label{fulll}
\dot{\rho}&=&-i[H_{\rm I},\rho]+\sum_{i=1}^N\frac{\gamma_r}{2}{\cal D}[|p\rangle_{ii}\langle r|]\rho+
\sum_{i=1}^N\frac{\gamma_r}{2}{\cal D}[|g\rangle_{ii}\langle r|]\rho
\nonumber\\&&
+\sum_{i=1}^N\gamma_p{\cal D}[|g\rangle_{ii}\langle p|]\rho+\kappa{\cal D}[U_{\rm fb}a]\rho,
\end{eqnarray}
where $H_{\rm I}$ is the full Hamiltonian of Eq.~(\ref{full}), and we have assumed the
branching ratios of the atomic decays from level $|r\rangle$ to $|p\rangle$ and  $|r\rangle$ to $|g\rangle$ are the same for simplicity. Generally speaking, the decay rate of the Rydberg state $\gamma_r$ is three orders of magnitude below the decay rate of the intermediate state $\gamma_p$. In Fig.~\ref{p5}(c), we plot the populations of quantum states in the process of producing the antisymmetric Bell state and tripartite DFS state, respectively. We see that under the joint actions of two decay rates $\gamma_r$ and $\gamma_p$, the system is stabilized into an entangled state mixed up by the ground state and all single excitation states. A large single photon detuning $\Delta_{a(b)}$ does substantially reduce the effect of spontaneous emission for $|p\rangle$, but it is at the cost of extending the convergent time and amplifying the influence of spontaneous emission for $|r\rangle$. Thence the tradeoff between $\gamma_p$ and $\gamma_r$ should be considered according to different parameters of system.

Now let us consider the basic elements that may be
candidates for the intended experiment. The cavity QED with Rydberg-blocked atoms is a favorable platform for implementing the current proposal \cite{Bre,diss,zxf,njp}. The transition between atomic ground level $5S_{1/2}$ and the optical level $5P_{3/2}$ of $^{87}$Rb atom is coupled to the quantized cavity mode with strength $g=2\pi\times14.4$ MHz.  The decay rates of the intermediate state $|p\rangle$, the Rydberg state $|r\rangle$ and the cavity mode are $\gamma_p=2\pi\times3$ MHz,  $\gamma_r=2\pi\times1$ kHz, and $\kappa=2\pi\times0.66$ MHz, respectively.
The Rabi laser frequency $\Omega_{c(B,R)}$ can be tuned continuously and we adopt $|\Omega_{c(B,R)}|=g$, and the single-photon detunings $\Delta_a$ and $\Delta_b$ are set to be $80g$ in order to preclude the excitation of the intermediate atomic state $|p\rangle$. By substituting these parameters into Eq.~(\ref{fulll}), the fidelities of bipartite and tripartite entanglement reach $98.31\%$ and $98.57\%$ for a short time $g_{\rm eff}t=20$, as demonstrated in Fig.~\ref{p5}(d).

In conclusion, we have shown that the quantum feedback control combined with Rydberg atoms can be exploited to
stabilize multipartite entanglement. The dimension of the system is effectively reduced due to the strong Rydberg blockade,
which is instrumental in simplifying the complexity of the quantum feedback control. Most interestingly, the entangled state is not restricted to a fixed form, but can be adjustable via tuning classical fields,  and a high fidelity is obtained with experimentally achievable parameters. We hope that this work may open a new venue for the experimental
realization of multipartite entanglement in the near future.
\begin{center}{\bf{ACKNOWLEDGMENT}}
\end{center}

This work is supported by Natural Science Foundation of China under Grant No. 11647308, No. 11547303, No. 11534002, and No. 61475033,
Fundamental Research Funds for the Central Universities under Grant No.  2412016KJ004.


\begin{thebibliography}{999}
\bibitem{sc}E. Schr\"{o}dinger, Proc. Cambridge Philos. Soc. {\bf 31}, 555 (1935).
\bibitem{ein} A. Einstein, B. Podolsky, and N. Rosen, Phys. Rev. {\bf47},
777 (1935).


\bibitem{pan1}D. Bouwmeester, J. W. Pan, M. Daniell, H. Weinfurter, and A. Zeilinger, Phys. Rev. Lett. {\bf82}, 1345 (1999).
\bibitem{panjw}J. W. Pan, M. Daniell, S. Gasparoni, G. Weihs, and A. Zeilinger, Phys. Rev. Lett. {\bf 86}, 4435 (2001).


\bibitem{ten}X. L. Wang, L. K. Chen, W. Li {\it et al}, arXiv:1605.08547 (2016).
\bibitem{haroche}A. Rauschenbeutel, G. Nogues, S. Osnaghi, P. Bertet, M. Brune, J. M. Raimond, S. Haroche, Science  {\bf 288}, 2024 (2000).
\bibitem{haroche1} J. M. Raimond, M. Brune, and S. Haroche, Rev. Mod. Phys. {\bf73}, 565 (2001).



\bibitem{ion1}Q. A. Turchette, C. S. Wood, B. E. King, C. J. Myatt, D. Leibfried, W. M. Itano, C. Monroe, and D. J. Wineland,  Phys. Rev. Lett. {\bf81}, 3631 (1998).
\bibitem{ion2} R. Blatt and D. J. Wineland, Nature (London) {\bf453}, 1008 (2008).
\bibitem{monz}T. Monz, P. Schindler, J. T. Barreiro {\it et al}, Phys. Rev. Lett. {\bf106}, 130506 (2011).


\bibitem{Bell} J. S. Bell, Physics (Long Island City, N.Y.) \textbf{1}, 195 (1965).



\bibitem{long}G. L. Long and X. S. Liu,
Phys. Rev. A {\bf65}, 032302 (2002).
\bibitem{deng}F. G. Deng and G. L. Long,
Phys. Rev. A {\bf68}, 042315 (2003).



\bibitem{ghz} D. M. Greenberger, M. A. Horne, and A. Zeilinger, 1989,
{\it Going beyond Bell¡¯s theorem in Bell¡¯s Theorem, Quantum
Theory, and Conceptions of the Universe} (Kluwer Academic,
Dorthecht)
\bibitem{wstate}W. D\"{u}r, G. Vidal, and J. I. Cirac,  Phys. Rev. A {\bf62}, 062314 (2000).





\bibitem{DFS1}D. A. Lidar, I. L. Chuang, and K. B. Whaley, Phys. Rev. Lett. {\bf81}, 2594 (1998).
\bibitem{duan}L. M. Duan and G. C. Guo, Phys. Rev. A {\bf58}, 3491 (1998).
\bibitem{DFS2}A. Beige, D. Braun, B. Tregenna, and P. L. Knight, Phys. Rev. Lett. {\bf85}, 1762 (2000).


\bibitem{eBell}A. Orieux, A. Eckstein, A. Lema\^{\i}tre, P. Filloux, I. Favero, G. Leo, T. Coudreau, A. Keller, P. Milman, and S. Ducci, Phys. Rev. Lett. {\bf110}, 160502 (2013).

\bibitem{ewstate}M. Neeley, R. C. Bialczak, M. Lenander, E. Lucero, M.
Mariantoni, A. D. O'Connell, D. Sank, H. Wang, M. Weides, J.
Wenner, Y. Yin, T. Yamamoto, A. N. Cleland, and J. M. Martinis,
Nature (London) {\bf467}, 570 (2010).

\bibitem{gw} D. R. Hamel, L. K. Shalm, H. H\"{u}bel, A. J. Miller, F. Marsili,
V. B. Verma, R. P. Mirin, S. W. Nam, K. J. Resch, and
T. Jennewein, Nat. Photon. {\bf8}, 801 (2014).

\bibitem{xiu}X. M. Xiu, Q. Y. Li, Y. F. Lin, H. K. Dong, L Dong, and Y. J. Gao,  Phys. Rev. A {\bf94},
042321 (2016).


\bibitem{zoller}T. Pellizzari, S. A. Gardiner, J. I. Cirac, and P. Zoller
Phys. Rev. Lett. {\bf75}, 3788 (1995).
\bibitem{zurek}W. H. Zurek, Rev. Mod. Phys. {\bf75}, 715 (2003).



\bibitem{wiseman} H. M. Wiseman, Phys. Rev. A {\bf49}, 2133 (1994).

\bibitem{mil} S. Schneider and G. J. Milburn, Phys. Rev. A {\bf65}, 042107 (2002).
\bibitem{wang}J. Wang, H. M. Wiseman, and G. J. Milburn, Phys. Rev. A {\bf71},
042309 (2005).


\bibitem{1}A. R. R. Carvalho and J. J. Hope, Phys. Rev. A {\bf76}, 010301(R)
(2007).
\bibitem{2}A. R. R. Carvalho, A. J. S. Reid, and J. J. Hope, Phys. Rev. A
{\bf78}, 012334 (2008).
\bibitem{3}R. N. Stevenson, J. J. Hope, and A. R. R. Carvalho,  Phys. Rev. A
{\bf84}, 022332 (2011).

\bibitem{shaoxq} X. Q. Shao, T. Y. Zheng, and S. Zhang, Phys. Rev. A
{\bf85}, 042308 (2012); X. Q.  Shao, Z. H. Wang, H. D. Liu, and X. X. Yi,
Phys. Rev. A {\bf94}, 032307 (2016).
\bibitem{ben}C. D. B. Bentley, A. R. R. Carvalho, D. Kielpinski, J. J. Hope, Phys. Rev. Lett. {\bf113}, 040501 (2014).


\bibitem{rd1} M. Saffman and K. M{\o}lmer, Phys. Rev. Lett.
{\bf102}, 240502 (2009).
\bibitem{rd2}M. Saffman, T. G. Walker, and K. M{\o}lmer, Rev. Mod. Phys. {\bf82}, 2313 (2010).
\bibitem{rd3}T. Wilk, A. Ga\"{e}tan, C. Evellin, J. Wolters, Y. Miroshnychenko, P.
Grangier, and A. Browaeys, Phys. Rev. Lett. {\bf104},
010502 (2010).
\bibitem{rd4}H. Z. Wu, Z. B. Yang, and S. B. Zheng, Phys. Rev. A 82, 034307
(2010).
\bibitem{rd5}M. M. M\"{u}ller, M. Murphy, S. Montangero, T. Calarco, P. Grangier, and A. Browaeys, Phys. Rev. A {\bf89}, 032334 (2014).
\bibitem{rd6}X. Q. Shao, J. B. You, T. Y. Zheng, C. H.
Oh, and S. Zhang, Phys. Rev. A {\bf89}, 052313 (2014).
\bibitem{rd7} M. Saffman, J. Phys. B {\bf49}, 202001 (2016).
\bibitem{scu}M. O. Scully and M. S. Zubairy, Quantum Optics (Cambridge
University Press, Cambridge, 1997).
\bibitem{Nielsen}M. A. Nielsen and I. L. Chuang, \emph{Quantum Computation and Quantum Information} (Cambridge
University Press, Cambridge, 2000).


\bibitem{Bre}F. Brennecke, T. Donner, S. Ritter, T. Bourdel, M. K\"{o}hl, and
T. Esslinger, Nature (London) 450, 268 (2007).
\bibitem{diss}C. Guerlin,  E. Brion, T. Esslinger, and K. M{\o}lmer, Phys. Rev. A {\bf82}, 053832 (2010).
\bibitem{zxf}X. F. Zhang, Q. Sun, Y. C. Wen, W. M. Liu, S. Eggert, and A. C. Ji, Phys. Rev. Lett. {\bf110}, 090402 (2013).
\bibitem{njp}A. Grankin, E. Brion, E. Bimbard, R. Boddeda, I. Usmani,
A. Ourjoumtsev, and P Grangier, New J. Phys. {\bf16}, 043020 (2014).
\end{thebibliography}
\end{document}